\documentclass[aps, prl, superscriptaddress, twocolumn, letterpaper, 10pt, amsfonts, amsmath, amssymb]{revtex4-2}

\usepackage{graphicx}
\usepackage{hyperref} 
\usepackage{xcolor}
\usepackage{soul}
\hypersetup{
    colorlinks,
    linkcolor={blue!80!black},
    citecolor={blue!80!black},
    urlcolor={blue!80!black}
}
\usepackage{physics}

\usepackage{placeins}

\usepackage{siunitx}
\usepackage{chemformula}

\usepackage{xcolor}
\usepackage{lipsum}

\definecolor{editcolor}{rgb}{1, 0.0, 0.0}

\begin{document}
\title{
Dissipationless Nonlinearity in Quantum Material Josephson Diodes}
\author{Constantin Schrade}
\affiliation{Center for Quantum Devices, Niels Bohr Institute, University of Copenhagen, 2100 Copenhagen, Denmark}
\author{Valla Fatemi}
\affiliation{School of Applied and Engineering Physics, Cornell University, Ithaca, NY, 14853, USA}

\date{\today}

\begin{abstract} 
Dissipationless nonlinearities for three-wave mixing are a key component of many superconducting quantum devices, such as amplifiers and bosonic qubits.
So far, such third-order nonlinearities have been primarily achieved with circuits of concatenated Josephson tunnel junctions. 
In this work, we theoretically develop an alternative approach to realize third-order nonlinearities from gate-tunable and intrinsically symmetry-broken quantum material Josephson junctions. 
We illustrate this approach on two examples, an Andreev interferometer and a magnetic Josephson junction. Our results show that both setups enable Kerr-free three-wave mixing for a broad range of frequencies, an attribute that is highly desirable for amplifier applications. Moreover, we also find that the magnetic junction constitutes a paradigmatic example for three-wave mixing in a minimal single-junction device without the need for any external biases.  
We hope that our work will guide the search of dissipationless nonlinearities in quantum material superconducting devices and inspire new ways of characterizing symmetry-breaking in quantum materials with microwave techniques. 
\end{abstract}

\maketitle

Recently, there has been a surge of interest in realizing non-reciprocal phenomena with superconducting quantum materials and devices~\cite{ando_observation_2020,daido2022intrinsic,he_phenomenological_2022,scammell_theory_2022,yuan_supercurrent_2022,legg2022superconducting}. 
A prominent example is the Josephson diode effect, which is characterized by a critical current that depends on the direction of an applied current-bias~\cite{zhang_general_2022,misaki_theory_2021,wu_field-free_2022,diez2023symmetry,hu2023josephson,zazunov2009anomalous,brunetti2013anomalous,baumgartner_supercurrent_2022,lotfizadeh2023superconducting,mazur2022gate,gupta2023gate,costa2023microscopic,zhang2022evidence,davydova_universal_2022,pal2022josephson,banerjee2023phase,zazunov2023nonreciprocal,zazunov2023approaching,souto2022josephson,kononov2020one,ciaccia2023gate,valentini2023radio,greco2023josephson,cuozzo2023microwave,legg2023parity,trahms2023diode,maiani2023nonsinusoidal,hess2023josephson}. 
To achieve this asymmetric current-bias response, a key requirement is the simultaneous breaking of inversion and time-reversal symmetry, which results in a non-symmetric current-phase relation.
Therefore, the Josephson diode effect has been regarded as novel indicator of such broken symmetries.

However, it is important to note that the Josephson diode effect requires biasing \textit{beyond} the supercurrent branch and, hence, requires dissipation. 
This constitutes a disadvantage for quantum information systems as dissipation leads to decoherence of quantum states. 
We motivate that the same broken symmetries also unlock the lowest-order nonreciprocal nonlinear response \textit{inside} the supercurrent branch: a dissipationless, nonreciprocal, nonlinear inductance. 
This nonlinearity, captured by the $\chi^{(2)}$ tensor in optics~\cite{boyd2008nonlinear} and as $g_3$ in resonant superconducting circuits~\footnote{We use the convention of N. Frattini, et al.~\cite{frattini_optimizing_2018}}, exhibits basic measurable effects like second-harmonic generation and parametric signal amplification by second-harmonic driving.
In superconducting quantum devices, such a Nonreciprocal Josephson dipole (NJD) device plays a pivotal role in many contexts, including parametric amplifiers~\cite{frattini_3-wave_2017,frattini_optimizing_2018,sivak_kerr-free_2019,miano_frequency-tunable_2022,frattini_three-wave_2021,khabipov2022superconducting,zorin2016josephson}, superconducting bosonic qubits~\cite{ma_quantum_2021,albert_bosonic_2022}, and nonlinear couplers~\cite{chapman_high_2022}.

\begin{figure}[!b]
    \centering
    \includegraphics[width=0.95\linewidth]{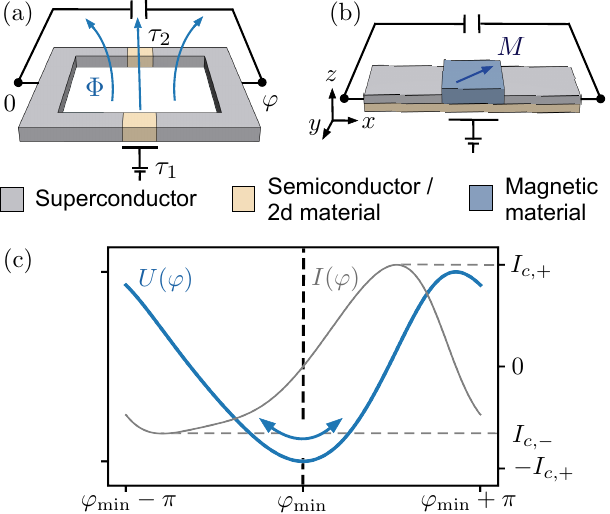}
    \caption{
    (a) Capacitively-shunted Andreev interferometer of two semiconductor or $2d$ material JJs, enabling a gate-tunable, Kerr-free, non-reciprocal Josephson dipole (NJD). 
    (b) Semiconductor or $2d$ material JJ proximitized by a magnet with magnetization, $M$, for a gate- and magnetization-tunable Kerr-free NJD.
    (c) Asymmetric Josephson potential (blue), $U(\varphi)$, realizing a $g_3$ nonlinearity for a NJD. The current-phase relation (gray), $I(\varphi)$, shows a diode effect, $I_{c,+}\neq |I_{c,-}|$ with $I_{c,\pm}$ the critical currents in the forward/reverse direction.
    }
    \label{fig:intro}
\end{figure}

So far, the necessary symmetry breaking for NJDs has been achieved by asymmetric concatenation of conventional Josephson tunnel junctions and magnetic fluxes (with Superconducting Nonlinear Asymmetric Inductive eLements, SNAILs, being a prominent example)~\cite{frattini_3-wave_2017,frattini_optimizing_2018,sivak_kerr-free_2019,miano_frequency-tunable_2022,frattini_three-wave_2021}. 
In contrast, in quantum material Josephson junctions (JJs), the relevant symmetry breaking can arise intrinsically, for example due to the magnetic material properties~\cite{wu_field-free_2022}. 
Thus, external magnetic fields during device operation can be avoided, opening the way to NJD devices with either no parametric bias whatsoever or fast control with gate voltages instead of magnetic flux. 
Such a NJD device can lead to simplified circuit designs or ones without current bias lines and, thereby, improved performance, density, or dissipation~\cite{larsen_semiconductor-nanowire-based_2015}. 
We thus argue that such NJDs are attractive targets for the quantum materials community.

In this work, we introduce and study two approaches towards a NJD element based on quantum material JJs. 
The first approach employs a superconducting interferometer with gate-tunable semiconductor or $2d$ material JJs.
The second approach involves a single JJ with a magnetic weak link. 
For both approaches, we show that a $g_3$ nonlinearity can be realized that can be tuned \textit{in-situ} with local gate electrodes. 
Notably, realizing the NJD effect in the single JJ approach does not necessitate the use of a strong external magnetic field. Instead, a fixed magnetization of an intrinsic magnetic component is sufficient.
Our findings also show that the $g_3$ nonlinearity can be attained in the absence of a Kerr effect, which is of high utility in devices like  
quantum-limited parametric amplifiers~\cite{sivak_kerr-free_2019,chapman_high_2022}. 
We hope that our work will inspire the future search for dissipationless nonlinearities in quantum material superconducting devices and, simultaneously, lead to novel application in quantum information platforms.

\begin{figure}[!b]
    \centering
    \includegraphics[width=\linewidth]{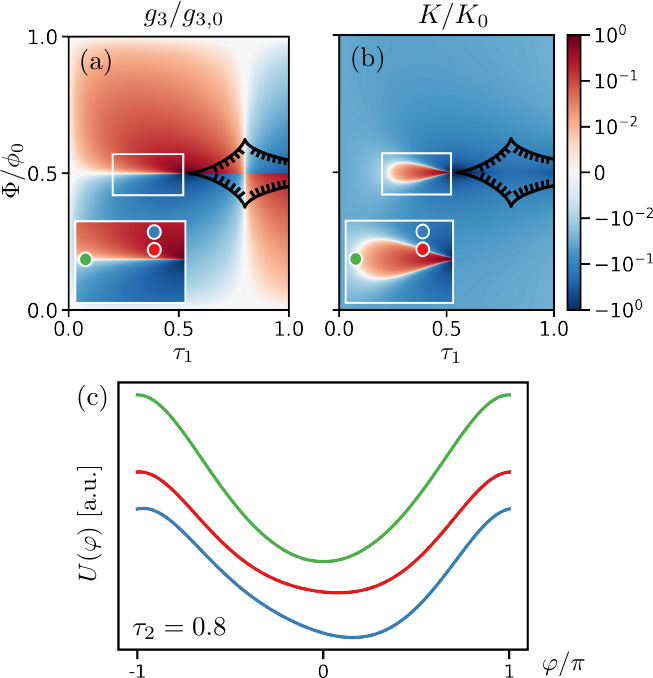}
    \caption{
    (a) Nonlinearity, $g_3$ with $\hbar g_{3,0} \equiv c_2 \varphi_{\text{zpf}}^3 /6$, for the Andreev interferometer versus flux, $\Phi$, and transmission, $\tau_{1}$ ($\tau_{2}=0.8$). In the region bounded by hashed lines, $U(\varphi)$ has two minima and $g_3$ was computed from the global minimum. 
(b) Same as (a) but for the Kerr coefficient, $K$ with $\hbar K_{0}\equiv c_2 \varphi_{\text{zpf}}^4 /2 $.
(c) Josephson potential, $U$, versus phase difference, $\varphi$, for various $(\tau_{1},\Phi)$, marked with colored dots in (a) and (b). Green curve (top) is mirror symmetric with $g_{3}=0$ and $K=0$. Remaining curves are non-symmetric with $g_{3}\neq 0$; red curve (center) has $K=0$.
    }
    \label{fig:2}
\end{figure}

\textit{Interferometer approach.} 
We begin with our first approach to a NJD: A capacitively-shunted superconducting quantum interference device (SQUID) comprised of two JJs, mediating the Josephson coupling via highly-transmissive Andreev bound states, see Fig.~\ref{fig:intro}(a). 
This structure is convenient since the construction of its energy-phase relations of short, high-transparency junctions are in close correspondence to a series of tunnel junctions~\cite{Bozkurt2023}, with the important benefit of gate-voltage tunability. 
Our goal is to describe how this setup realizes a NJD with a Kerr-free $g_3$ nonlinearity and permits fast control by gate voltages instead of magnetic fluxes.

As a first step, we introduce the Hamiltonian of our interferometer setup, 
\begin{equation}
\hat{\mathcal{H}}
=4E_{C}\hat{N}^{2}
+
U_{1}(\hat\varphi)
+
U_{2}(\hat\varphi- \Phi/\phi_0).
\label{Hfull}
\end{equation}
In the first term, $E_{C}=e^{2}/2C$ describes the charging energy, where $C$ is the shunt capacitance and the operator $\hat{N}$ counts the number of Cooper-pairs that have passed through the interferometer~\cite{vool_introduction_2017}. 
$\hat{N}$ is conjugate to the superconducting phase operator, $\hat\varphi$.
The subsequent terms, $U(\varphi)\equiv U_{1}(\varphi)+U_{2}(\varphi - \Phi/\phi_0)$, describe the Josephson potential. 
$\phi_{0}=\hbar/2e$ is the reduced flux quantum, and $U_1(\varphi)$ and $U_2(\varphi)$ are the ground-state energy-phase relations of the Andreev bound states in the two junctions. 
In the short-junction limit, these energy-phase relations take on the form~\cite{beenakker1991universal,sauls2018andreev}, 
\begin{equation}
U_{\ell}(\varphi)    
=
-\Delta\sqrt{1-\tau_{\ell}\sin^{2}(\varphi/2)}, \label{Ephi1}
\end{equation}
where $\Delta$ is the superconducting gap and $\tau_{\ell}$ is the gate-tunable transmission of the Andreev bound state in the $\ell^{\text{th}}$ junction. 
For low transmissions ($\tau_{\ell}\ll 1$), Eq.\,\eqref{Ephi1} is dominated by the first harmonic, $U_{\ell}(\varphi)\approx -U_{\ell,1}\cos\varphi$. 
For high transmissions ($\tau_{\ell}\lesssim 1$), higher harmonics contribute, $U_{\ell}(\varphi)= -U_{\ell,1}\cos\varphi-U_{\ell,2}\cos2\varphi+\dots$. 
These higher harmonics distinguish junctions hosting low lying Andreev bound states from single conventional Josephson tunnel junctions and will be essential for realizing nonreciprocal effects.

To investigate these nonreciprocal effects, we now turn our attention to the limit of a large shunting capacitance so that zero-point fluctuations (zpf) of the superconducting phase are small, $\varphi_{\text{zpf}}\equiv(2E_{C}/c_2)^{1/4}\ll1$ where $c_2$ is the second-order Taylor coefficient of $U(\varphi)=\sum\nolimits_{j=2}^{\infty}(c_{j}/j!)(\varphi-\varphi_{\text{min}})^{j}$.
The number and phase operators are then conveniently expressed in terms of bosonic raising and lowering operators, $\hat\varphi=\varphi_{\text{zpf}}(\hat{a}+\hat{a}^{\dag})$ and $\hat{N}=i(1/2\varphi_{\text{zpf}})(\hat{a}-\hat{a}^{\dag})$, which permit the expansion of the Hamiltonian in Eq.\,\eqref{Hfull} around the global minimum, $\varphi_{\text{min}}$, of $U(\varphi)$.
In terms of these bosonic operators, $\hat{\mathcal{H}}$ takes on the form of a weakly-nonlinear oscillator Hamiltonian, 
\begin{equation}
    \hat{\mathcal{H}} = \hbar \omega_r \hat{a}^\dagger \hat{a} + 
    \sum_{m\geq 3} \hbar g_{m} (\hat{a} + \hat{a}^\dagger)^m ,
    \label{expansion}
\end{equation}
where $\omega_{r}=1/\sqrt{LC}=\sqrt{8E_C c_2/\hbar^2}$ is the resonator frequency with the inductance $L^{-1}=\phi_0^{-2} c_2$~\cite{frattini_optimizing_2018}. 
$g_m$ denotes the $m^{\text{th}}$-order nonlinearity at the single-photon level~\footnote{An alternative approach to finding the perturbative expansion at the Hamiltonian extrema is given in~\cite{miano2023hamiltonian}.}.

The third-order nonlinearity, $g_3$, in Eq.\,\eqref{expansion} is the key characteristic of non-reciprocity at the single-photon level. 
From the described expansion procedure of the Hamiltonian, $g_3$ is obtained as,
\begin{equation}
    \hbar g_3 =\frac{c_3}{6}\varphi_{\text{zpf}}^{3},
    \label{g3_expr}
\end{equation}
where $c_3$ is the third-order Taylor coefficient of $U(\varphi)$~\cite{frattini_optimizing_2018}. From Eq.\,\eqref{g3_expr}, we see that a finite third-order nonlinearity, $g_{3}\neq 0$, is induced by a Josephson potential $U(\varphi)$ that is not mirror-symmetric around $\varphi_{\text{min}}$ so that $c_{3}\neq0$.

\begin{figure}[!b]
    \centering
    \includegraphics[width=\linewidth]{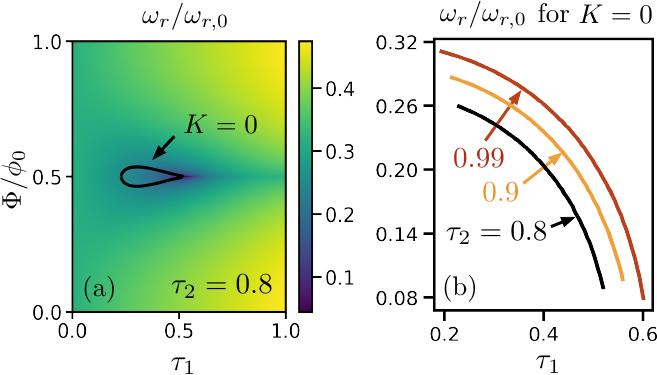}
    \caption{
    (a) Resonator frequency, $\omega_r$ with $\omega_{r,0}=\sqrt{\Delta/C}/\phi_0$, versus the flux, $\Phi$, and the transmission, $\tau_1$ ($\tau_{2}=0.8$). Solid black lines indicate a vanishing Kerr coefficient, $K(\Phi,\tau_1)=0$. (b) Resonator frequencies, $\omega_r$, versus the transmission, $\tau_{1}$, along the $K(\Phi(\tau_{1}),\tau_1)=0$ arc in (a). Colored lines correspond to different values of $\tau_{2}$. For any $\tau_{1}$ in the plot, we pick the unique $\Phi(\tau_{1})>\phi_{0}/2$ at which $K(\Phi(\tau_{1}),\tau_1)=0$. 
    }
    \label{fig:3}
\end{figure}

Besides the third-order nonlinearity, $g_3$, it is important to emphasize that also the higher nonlinearities in Eq.\,\eqref{expansion} play an essential role. 
Most critically, the fourth-order nonlinearity, $g_4$, contributes to the Kerr effect, which corresponds to a contribution $\hbar K (\hat{a}^{\dag}\hat{a})^{2}$ in $\hat{\mathcal{H}}$ with the Kerr coefficient $K$. 
These Kerr terms give rise to an amplitude-dependent frequency shift of the oscillator and, from an application perspective, limit the maximum signal power in parameteric amplifiers, one of the chief limitations of such devices~\cite{frattini_3-wave_2017,frattini_optimizing_2018,sivak_kerr-free_2019,miano_frequency-tunable_2022,frattini_three-wave_2021}. 
Hence, minimizing the Kerr term is often a target for applied quantum devices.  
The Kerr term can be computed as,
\begin{equation}
    \hbar K = \frac{1}{2}\left(c_4 -  \frac{5c_3^2}{3c_2}\right) \varphi_{\text{zpf}}^{4}, 
    \label{K_expr}
\end{equation}
with $c_4$ the fourth-order Taylor coefficient of $U(\varphi)$~\cite{frattini_optimizing_2018}. 
In the following, we will use Eqs.\,\eqref{g3_expr} and \eqref{K_expr} to show how the third-order nonlinearity, $g_3$, and the Kerr coefficient, $K$, can be controlled by means of the gate-tunable transmissions, $\tau_{\ell}$, and the magnetic flux, $\Phi$.

Starting with the third-order nonlinearity, our expectation is that it will arise if time-reversal symmetry is broken by the magnetic flux, $\Phi/\phi_0 \neq n \pi$ for integer $n$, and inversion symmetry of the interferometer loop is \textit{simultaneously} broken by unequal transmission, $\tau_1 \neq \tau_2$. 
To explore this hypothesis, we have computed $g_3$ versus $(\Phi,\tau_1)$ for fixed $\tau_2$. 
Our results are shown in Fig.~\ref{fig:2}(a). We find that $g_3$ takes on nonzero values within four distinct lobes of the $(\Phi,\tau_{1})$-phase diagram separated by the symmetry-restoring conditions, 
$\Phi/\phi_{0}=n\pi$ or $\tau_{1}=\tau_{2}$. 
Upon transitioning through the symmetry-restoring conditions, we find that $g_3$ changes sign and vanishes exactly when time-reversal \textit{or} inversion symmetry are restored. 
In addition, if \textit{both} $\tau_{\ell}\ll 1$, we find that $g_{3}$ is suppressed \textit{independently} of the symmetries of the setup~\cite{Supplemental}. 
This suppression arises because higher harmonics are suppressed and the Josephson potential, $U(\varphi)\approx -U_{1}\cos(\varphi-\varphi_0)$ for some phase shift $\varphi_0$, is always symmetric around its minimum. 
This finding highlights the importance of higher harmonics in the Josephson potential for achieving a finite $g_3$.

It now natural to ask how these results for $g_3$ correlate with the Kerr coefficient, $K$. 
The computation of $K$ for the same parameter set as $g_3$ is shown in Fig.~\ref{fig:2}(b). 
We find that $K$ vanishes along a closed contour in the $(\Phi,\tau_{1})$-phase diagram, while $g_{3}$ remains finite for the same parameter values. 
Notably, this observation implies that the Andreev interferometer can realize a gate- and flux-tunable Kerr-free third-order nonlinearity. 
In addition, we also find that the resonator frequency, $\omega_r$, varies along the $K=0$ arc, per Fig.~\ref{fig:3}. 
Consequently, our setup enables tunable resonator frequency hosting $g_3 \neq 0 $ while maintaining the desirable $K=0$ condition. 
Such an attractive feature was, so far, only found within circuits that involve multiple flux-bias lines~\cite{miano_frequency-tunable_2022}. 
In contrast, our approach only uses a single flux-bias, and the dissimilar tuning knobs (one voltage, one flux) have no cross-couplings at DC. 

Finally, we highlight a distinctive feature of the $(\Phi,\tau_{1})$-phase diagrams of Fig.~\ref{fig:2}, which is a concurrent suppression of $g_{3}$ and $K$ when the $K=0$ arc intersects $\Phi=0.5\phi_0$. 
At this ``sweet spot", a sixth-order term is the lowest-order nonlinearity.
As pointed out in earlier work~\cite{mundhada2019experimental,miano2023hamiltonian}, such a higher-order nonlinarity could form an important component of hardware-efficient quantum error correction based on continuous-variable error-correction codes.

\begin{figure*}[!t]
    \centering
    \includegraphics[width=1.0\textwidth]{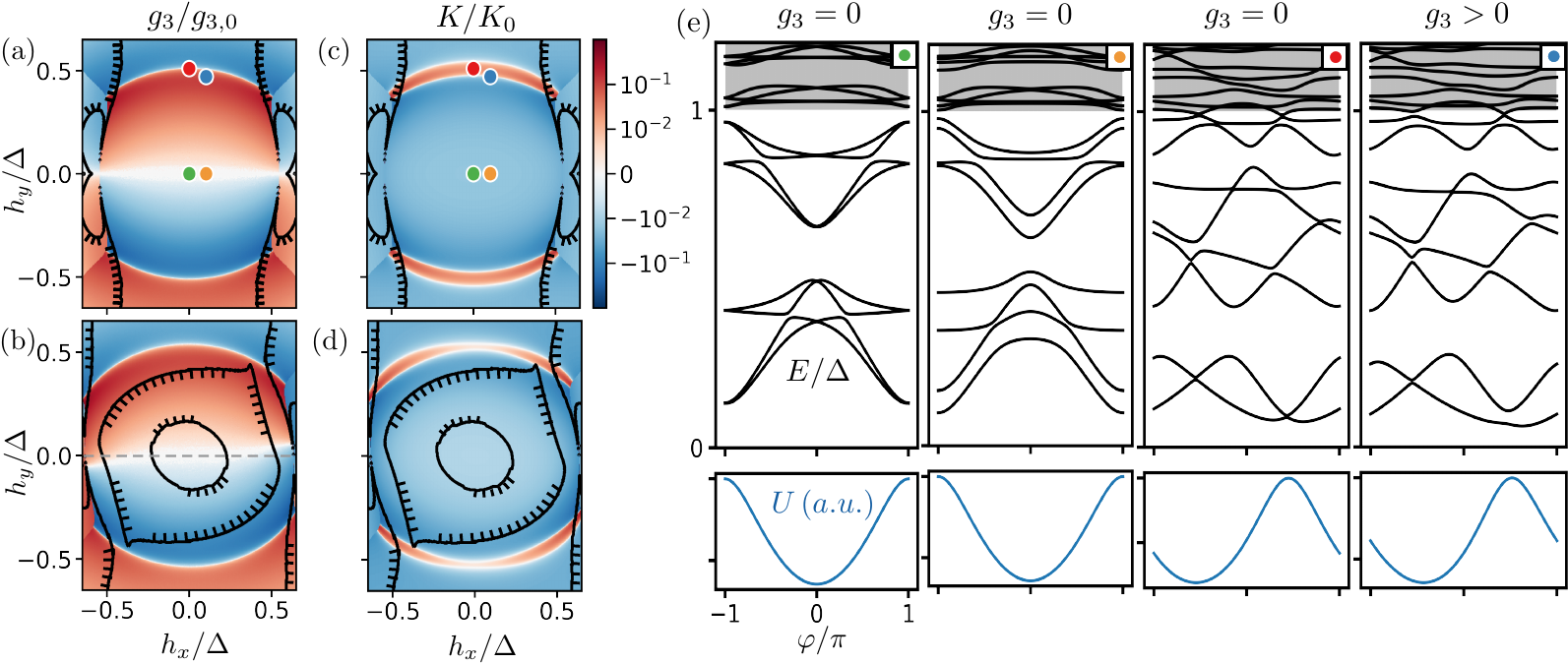}
    \caption{
    (a) Third-order nonlinearity, $g_3$, for a magnetic JJ versus the exchange splittings, $h_{x}$ and $h_{y}$. Detailed simulation parameters are given in the Supplemental Material~\cite{Supplemental}. (b) Same as (a) but for the case when potential impurities break the mirror symmetry. In this case, $g_{3}\neq0$ even if $h_{y}=0$. (c-d) Same as (a-b) but for the Kerr coefficient, $K$.
(e) Andreev bound state spectra and Josephson energies, $U(\varphi)$, versus superconducting phase, $\varphi$, for points marked in (a-b). If $h_{y}=0$, there is no nonreciprocity. If $h_{y}\neq0$, $U(\varphi)$ has no mirror symmetry except at special points where $g_{3}$ and $\eta$ change sign, as marked with a red dot in (a-b). 
The discreteness of the levels above $\Delta$ arise from the finite size of the model. 
    }
    \label{fig:1junction}
\end{figure*}

\textit{Single junction approach.}
We now turn to a second approach for realizing a NJD: a single JJ with a magnetic weak link, see Fig.~\ref{fig:intro}(b). 
Our goal is to show that this setup also implements a NJD with $g_3$ nonlinearity that does not require any external magnetic fields.

For concreteness, we will focus our considerations on a few-channel semiconductor JJ setup. 
In this case, inversion-symmetry breaking arises from a Rashba spin-orbit coupling that is either intrinsic to the semiconductor or due to interfacial electric fields. 
Meanwhile, time-reversal symmetry breaking is realized by magnetism locally applied to the weak link channels, for example due to a proximitizing magnetic material~\cite{liu2019semiconductor,razmadze2023supercurrent,wei2015strong,jiang2015independent,lee2016direct} or due to detached micromagnets~\cite{yoneda2015robust}. 
Such semiconductor JJs have shown a diode effect when subjected to an external magnetic field, indicating that the correct symmetry-breaking is present~\cite{baumgartner_supercurrent_2022,lotfizadeh2023superconducting,mazur2022gate}.

To model the setup, we employ a tight-binding simulation that describes the semiconductor JJ in terms of two transversal conduction channels~\cite{groth2014kwant}. 
This approach has been used for modelling experimentally observed Andreev bound state spectra in such JJ devices and other mesoscopic superconducting systems~\cite{tosi2019spin,hays2021coherent,matute2022signatures}. 
The effective Zeeman field components within the junction regions are denoted by $h_{x}$, for the $x$-direction perpendicular to the junction, and $h_y$, for the $y$-direction parallel to the junction~\footnote{For convenience, we use the symbol $h$ that is often used for exchange fields~\cite{strambini2017revealing}}. 
The effective electric field that produces the Rashba spin-orbit couplings is assumed here to point out of the junction plane in the $z$-direction. 
Further details on the tight-binding simulation are given in the Supplemental Material~\cite{Supplemental}.
The results of our simulations are shown in Fig.~\ref{fig:1junction}. 

Our main finding is that the $g_3$ nonlinearity is typically nonzero and determined by both the magnitude and orientation of the effective Zeeman field in the junction region, see Fig.~\ref{fig:1junction}(a). 
There are two notable aspects of the dependence on the effective Zeeman field:

(1) When the effective Zeeman field points along the junction ($h_{y}=0$), the Andreev bound state spectrum and the Josephson energy are symmetric around $\varphi=0$, as shown in Fig.~\ref{fig:1junction}(e).
As a result, no nonreciprocity is induced (i.e., $g_3=0$), despite the violation of time-reversal and inversion symmetries. The reason for vanishing $g_3$ is that the effective Zeeman field along the junction, $h_x$, does not break the spatial mirror symmetry across the $y$-$z$ plane, nor does the electric field inducing the spin-orbit interaction, which aligns with the $z$-direction. 

To further emphasize the importance of this mirror symmetry breaking, we introduced potential impurities that break mirror symmetry in the junction region. As shown in Fig.~\ref{fig:1junction}(b), we find that now $h_x$ alone is sufficient to generate a finite $g_3$, although it is weak without nonzero $h_y$. In the Supplemental Material~\cite{Supplemental}, we also demonstrate similar symmetry properties in the diode efficiency.   Thus, we identify mirror symmetry breaking as a novel approach to achieve a nonreciprocal $g_3$ nonlinearity for arbitrary magnetizations. Furthermore, since the emergence of $g_3$ does not require precise tuning of the chemical potential in the junction, our proposed setup has the potential to realize a NJD device with no parametric bias whatsoever.

(2) Besides the universal emergence of nonreciprocal nonlinearity, another important feature of $g_3$ are its sign reversals at finite values of ($h_x, h_y$), see Fig.~\ref{fig:1junction}(a-b). In the Supplemental Material~\cite{Supplemental}, we show that the critical current asymmetry, $\Delta I_{c}=I_{c,+}-|I_{c,-}|$, which is the key characteristic of the Josephson diode effect, exhibits similar sign reversal, consistent with recent experimental findings~\cite{lotfizadeh2023superconducting}. Interestingly, the Kerr coefficient, $K$, also undergoes sign changes (Fig.~\ref{fig:1junction}(c-d)). 
These sign reversals occur at ($h_x, h_y$) values that are generally different from those resulting in sign reversals of $g_3$. Thus, the magnetized JJ constitutes a paradigmatic example for achieving a Kerr-free third-order nonlinearity in a minimal single-junction device.

\textit{Conclusion.} 
In this work, we explored theoretically the dissipationless nonreciprocal nonlinearities of Josephson devices, Andreev interferometers and magnetic JJs. 
For both Josephson devices, we have identified strong third-order nonlinearities for a wide range of parameters, motivating new applications for such devices as gate-tunable or bias-free nonreciprocal nonlinear quantum device that are useful for 
parametric amplifiers~\cite{frattini_3-wave_2017,frattini_optimizing_2018,sivak_kerr-free_2019,miano_frequency-tunable_2022,frattini_three-wave_2021,khabipov2022superconducting,zorin2016josephson}, Kerr-cat qubits~\cite{grimm_stabilization_2020,frattini_squeezed_2022,venkatraman_driven_2023}, and nonlinear couplers~\cite{chapman_high_2022}.

\textit{Acknowledgements.} 
We acknowledge helpful discussions with Carlo Ciaccia, Reinhold Egger, Nicholas E. Frattini, Neda Foroozani, Silas Hoffman, Charles M. Marcus, Alessandro Miano, Jesper Nygard, Christian Sch\"onenberger, and Alex Zazunov.
We also acknowledge helpful comments from and related collaborations with Anton Akhmerov, Mert Bozkurt, and Morten Kjaergaard. 
CS thanks Peter D. Johannsen for ongoing collaborations on related works. CS acknowledges support from the Microsoft Corporation.

\textit{Data availability.} The code used to generate the figures of the main text is available on Zenodo~\cite{zenodolink}. 

\bibliographystyle{apsrev4-2}

\begin{widetext}

\begin{center}
\large{\bf Supplemental Material to `Dissipationless Nonlinearity in Quantum Material Josephson Diodes' \\}
\end{center}
\begin{center}
Constantin Schrade$^{1}$ and Valla Fatemi$^{2}$
\\
{\it $^{1}$Center for Quantum Devices, Niels Bohr Institute, University of Copenhagen, 2100 Copenhagen, Denmark}
\\
{\it $^{2}$School of Applied and Engineering Physics, Cornell University, Ithaca, NY, 14853, USA}
\end{center}
In the Supplemental Material, we provide more details and results on the interferometer and single-junction approach for realizing a nonreciprocal Josephson dipole (NJD). 
\end{widetext}

\setcounter{figure}{0}
\setcounter{equation}{0}
\renewcommand\thefigure{S\arabic{figure}}
\newpage

\section{Interferometer approach}
In this first section of the Supplemental Material, we provide details and results on the interferometer approach for realizing the NJD.

\subsection{Tunability with local gates}
In the main text, we have focused on analyzing the nonlinearity, $g_3$, and the Kerr coefficient, $K$, as a function of the transmission, $\tau_{1}$, and the magnetic flux, $\Phi$. 
Specifically, for the results presented in Fig.\,2 of the main text, the transmission of the second Andreev JJ was held constant at $\tau_{2}=0.8$. 
It is natural to ask, how our obtained results change upon varying $\tau_{2}$.

To address this question, we have computed $g_{3}$, $K$, and the signed diode efficiency $\eta=\Delta I_{c}/(I_{c,+}+|I_{c,-}|)$ with $\Delta I_{c}\equiv I_{c,+}-|I_{c,-}|$ for various values of $\tau_{2}$. Our results are depicted in Fig.\,\ref{fig:FigSM1}. Our primary finding is that $g_3$ remains finite within the same four distinct lobes, separated by the symmetry-restoring conditions, as outlined in detail in the main text. However, for the Kerr coefficient, $K$, we find that the arc length of the $K=0$ condition enlarges with larger $\tau_{2}$. Hence, we conclude that an increased $\tau_{2}$ offers a wider range of $\tau_{1}$ and $\Phi$ values to achieve the desirable $K=0$ condition.

An alternative approach for assessing the gate-tunability of $g_3$ and $K$ is to fix the magnetic flux, $\Phi$, and vary both transmissions, $\tau_{1}$ and $\tau_{2}$. Our results for this scenario are shown in Fig.\,\ref{fig:FigSM2}(a) and (b) with the magnetic flux being fixed to $\Phi=0.52\phi_0$. We find that $g_3$ is, in general, finite and vanishes along the $\tau_{1}=\tau_{2}$ diagonal at which inversion symmetry is restored. For the Kerr coefficient, we again identify a sizable arc along which the $K=0$ condition is satisfied. As depicted in Fig.\,\ref{fig:FigSM2}(c) and (d), the resonator frequency, $\omega_{r}$, can be adjusted over a broading range while maintaining the $K=0$ condition. This finding implies that a Kerr-free $g_3$ nonlinearity, tuned solely by local gates, can be achieved over a broad range of resonator frequencies.

\subsection{Double-minimum Josephson energy}
In the main text, our considerations focused on the parameter regimes when the Josephson energy only exhibits a single minimum. However, we indicated regions of parameter space that exhibit a double-minimum structure. We will now discuss the specifics of how this double-minimum Josephson energy behaves upon varying of the magnetic flux, $\Phi$, and the transmissions, $\tau_{1}$ and $\tau_{2}$.

Initially, we consider the case when $\Phi=0.5\phi_{0}$ and $\tau_{1}=\tau_{2}$, corresponding to the central point in the four-lobe structure of the $(\Phi,\tau_{1})$ phase diagram, shown Fig.\,\ref{fig:FigSM2}(e). At this parameter point, the Josephson energy comprises two local minima of equal depth. Such a Josephson energy has been utilized previously for a protected superconducting qubit that encodes the qubit degrees of freedom encoded in states of opposite Cooper-pair parity ~\cite{larsen2020parity2,schrade2022protected2,maiani2022entangling2,ciaccia2023charge2,leroux2023cat2}. Such an approach to error-protected quantum computing is complementary 
to approaches based on Majorana zero modes~\cite{hoffman2016universal2,landau2016towards2,plugge2017majorana2,schrade2018majorana2,schrade2022quantum2}. 

Next, we consider the situation marked by the green circle in Fig.\,\ref{fig:FigSM2}(e). At this parameter point, the transmissions are detuned such that $\tau_{1}\neq\tau_{2}$, but $\Phi=0.5\phi_{0}$ is maintained. This detuning of the transmissions changes the barrier height between the two minima of the Josephson energy, but the depth of the two minima remains the same. As noted in the caption of Fig.\,\ref{fig:FigSM2}(e), we have always used the minimum within $\varphi\in[0,\pi]$ to calculate the $g_3$ nonlinearity under such conditions.

Lastly, we consider the case of the blue circle in Fig.\,\ref{fig:FigSM2}(e), where we have slightly detuned the magnetic flux $\Phi\neq0.5\phi_0$ but retained $\tau_{1}=\tau_{2}$. In this situation. the relative depth of the two minima changes and we have used the deeper minima to compute $g_{3}$. In the more general case, when also $\tau_{1}\neq\tau_{2}$, corresponding to the red circle in Fig.\,\ref{fig:FigSM2}(e), the Josephson energy can be interpreted as arising from a combined detuning of the depth of the double-minima and a change in the barrier height between the minima.

\subsection{Sixth-order nonlinearity}
In the main text, we have highlighted a scenario for which both $g_{3}=0$ and $K=0$. 
This situation corresponds to the parameter point in the $(\Phi,\tau_{1})$ phase diagram of Fig.\,2 when the $K=0$ arc intersects $\Phi=0.5\phi_0$. At this specific parameter point, the leading order nonlinearity is sixth-order and obtained from the expansion of $\hat{\mathcal{H}}$ as~\cite{frattini_three-wave_20212},
\begin{equation}
\hbar g_{6}=\frac{1}{30}\frac{c_6}{c_2}\left(
\frac{E_C}{\hbar\omega_r}
\right)^{2}.
\end{equation}

To illustrate the emergence of this leading sixth-order nonlinearity, we have computed 
$g_3$ and $K$ as a function of $\tau_{1}$ for fixed $\Phi=0.5\phi_0$ and $\tau_{2}=0.8$, see Fig.\,\ref{fig:FigSM3}(a) and (b). 
The line plots in these figures allow us to pinpoint the transmission $\tau_{1}$ at which both $g_{3}=0$ and $K=0$, highlighted by a black arrow in Fig.\,\ref{fig:FigSM3}(b). We then contrast the behaviors of $g_3$ and $K$ to that of the sixth-order nonlinearity, $g_6$. Importantly, our findings in Fig.\,\ref{fig:FigSM3}(c) show that the sixth-order nonlinearity remains finite at the transmission where $g_3$ and $K$ vanish. This observation implies that the interferometer approach can indeed implement an unusual leading sixth-order nonlinearity.

\section{Single junction approach}
 In this second section of the Supplemental Material, we provide details and results on the single junction approach for realizing the NJD. 
\subsection{Tight-binding simulation}
We begin by discussing in detail our the tight-binding simulation, which was used to obtain the results of Fig.\,4 in the main text. 
The Hamiltonian for this tight-binding simulation of a semiconductor Josephson junction comprises, in a minimal approach, two transversal channels ($j=1,2$). In the absence of the magnetic component, it is given by, 
\begin{equation}
\begin{split}
H_{0}
&=
\sum_{i,j,s}(\varepsilon_{ij}-\mu_{i})\,c^{\dagger}_{i,j,s}c_{i,j,s}
+
t_{x,i}\,c^{\dagger}_{i+1,j,s}c_{i,j,s}
\\
&
+
s
\alpha_{x}\,c^{\dagger}_{i+1,j,s}c_{i,j,\bar{s}}
+
\sum_{i,j}\Delta_{i}e^{i\varphi_{i}}\,c_{i,j,\downarrow}c_{i,j,\uparrow}
\\
&+\sum_{i,s}t_{y}\,c^{\dagger}_{i,1,s}c_{i,2,s}
+
i\alpha_{y}\,c^{\dagger}_{i,1,s}c_{i,2,\bar{s}}+
\text{H.c.}
\end{split}
\label{Eq1}
\end{equation}
Here, $c_{i,j,s}$ is the electron annihilation operator with spin $s=\uparrow,\downarrow$ for a longitudinal site $i$ and transversal site $j$. 
The on-site energies are given by $\varepsilon_{ij}$ and measured relative to the chemical potential $\mu_{i}$. The hoppings in the longitudinal and transversal directions are denoted by $(t_{x,i},t_{y})$ and, correspondingly, the strength of spin-orbit couplings in those directions are denoted by $\alpha_{x,y}$. 
The magnitude and phase of the superconducting order parameter are $\Delta_{i}$ and $\varphi_{i}$. 
For simplicity, we will for now assume that the magnitude of the superconducting order parameter is constant and identical in both superconducting leads, $\Delta_{i}=\Delta$ for any superconducting site $(i,j)$. 
For the superconducting phase, we set $\varphi_{i}=\varphi$ if the site $(i,j)$ is in the left superconducting lead and $\varphi_{i}=0$ otherwise. 

In terms of the system parameters, we choose a set similar to the one used for modeling the experimental data in~\cite{matute2022signatures2}: $N_{\text{N}}=7$ and $N_{\text{SC}}=10$ for the number of lattice sites in the normal (N) and superconducting (SC) regions.
The lattice constants are taken as $a_{x}=50\,\text{nm}$ and $a_{y}=100\,\text{nm}$, and the superconducting gap magnitude is chosen to be $\Delta=0.2\,\text{meV}$. 
For the hopping amplitudes in the longitudinal direction, we choose for the normal and superconducting regions $(t_{x,\text{N}},t_{x,\text{S}})=(-0.85,-1)\cdot t_{0}/a^{2}_{x}$ with $t_{0}=\hbar^{2}/2m^{*}$. 
In the expression for $t_0$, we have introduced the effective electron mass in an Indium arsenide semiconductor, $m^{*}=0.023m_{e}$ with $m_e$ being the bare electron mass. 
For the hopping amplitudes in the transversal direction, we choose $t_{y}=-t_{0}/a_{y}^{2}$.
For the spin-orbit coupling, we pick $\alpha_{x}=\alpha/(2a_{x})$ and $\alpha_{y}=\alpha/(2a_{y})$ with $\alpha=15.5\,\text{meV}$. 
The on-site energies in the superconducting and the normal regions are set to $\varepsilon_{ij,\text{S}}=(2t_{0}/a^{2}_{x})-\Delta$ and $(\varepsilon_{i1,\text{N}},\varepsilon_{i2,\text{N}})=(1.2,1.08)\cdot 2t_{0}/a_{x}^{2}$, respectively. 
Lastly, we set $\mu_{i,\text{S}}=0$ for the chemical potential in the superconducting regions and $\mu_{i,\text{N}}=0.52\,\text{meV}$ for the normal regions.

As a next step, we introduce the Hamiltonian of the magnetic component, 
\begin{equation}
\begin{split}
H_{h}    
&=
\sum_{i,j,s,s'}(h_{x,i}\,\sigma^{x}_{ss'}+h_{y,i}\,\sigma^{y}_{ss'})\, c^{\dagger}_{i,j,s}c_{i,j,s'}
\\
&+
\sum_{i,j,s,s'}h_{z,i}\,\sigma^{z}_{ss'}\, c^{\dagger}_{i,j,s}c_{i,j,s'}.
\end{split}
\end{equation}
Here, $(h_{x,i},h_{y,i},h_{z,i})$ denote the exchange splittings due to the magnetization components in the $x$, $y$, and $z$-directions. The corresponding spin-space Pauli matrices are denoted by $\sigma^{x,y,z}$. We note that the exchange splittings are assumed to be non-zero only in the normal region. Moreover, in the main text, we also assume that the magnetization is in the junction plane, so that $h_{z,i}=0$.

In summary, the full Hamiltonian for our tight-binding simulations is given by $H_{0}+H_{h}$, which was used to obtain the results presented in Fig.\,4 of the main text. 

\subsection{Numerical procedure for obtaining the Taylor coefficients of the Josephson energy} 

For the interferometer approach, the Taylor coefficients, $c_m$, of the Josephson energy, $U(\varphi)$ can be obtained analytically from the derivatives of the expression in Eq.\,(2) of the main text. Such an analytical approach is not possible for the single-junction approach defined by the tight-binding Hamiltonian, $H_{0}+H_{h}$. For that reason, we need to use a numerical procedure for obtain the Taylor coefficients of the Josephson energy, which are given by, 
\begin{equation}
c_{m}=\frac{1}{m!}\partial_{\varphi}U(\varphi_{\text{min}}),
\end{equation}
where $\varphi_{\text{min}}$ is the superconducting phase difference at the global minimum of $U(\varphi)$.

To numerically obtain the Taylor coefficients, $c_m$, we follow the following three-step procedure: In a first step, we identify the superconducting phase difference, $\varphi_{\text{min}}$, at the global minimum of the Josephson energy. In a second step, we evaluate $U(\varphi)$ within an interval $[\varphi_{\text{min}}-\varepsilon,\varphi_{\text{min}}+\varepsilon]$ around this minimum phase for a fixed number, $N_{d}$, of equally spaced superconducting phase values. In a final step, we employ a least squares approach to fit a polynomial,
\begin{equation}
\sum_{m=2}^{M}c_{m}(\varphi-\varphi_{\text{min}})^{m},
\end{equation}
to the calculated values of $U(\varphi)$ within the defined interval. This fit gives the approximate values for the Taylor coefficients, $c_m$. For the results presented in this work, we typically used $\varepsilon=0.25$ for the interval length, $N_{d}=100$ for the discretization of the interval, and $M=10$ for the degree of the polynomial.

\subsection{Tunability with a local gate}
In the main text, we have computed the nonlinearity, $g_3$, and the Kerr coefficient, $K$, in Fig.\,4, keeping the chemical potential constant and varying the exchange splittings $(h_{x},h_{y})$. Specifically, we identified a contour where $K=0$ and $|g_{3}|>0$. An immediate question arises if the Kerr-free nonlinearity can be adjusted for different chemical potentials within the junction region, which are tunable by a local gate voltage.

To answer this question, we evaluated $g_3$ and $K$ as a function of the exchange field, $h_{y}$, and the chemical potential, $\mu$, as shown in Fig.\,\ref{fig:FigSM4}. Our key finding is that the Kerr-free condition for the $g_3$ nonlinearity can be retained across a range of chemical potentials, $\mu$, as illustrated in Fig.\,\ref{fig:FigSM4}(a) and (b). The normal state spectrum and the Andreev excitation spectrum are presented for selected values of $(h_{y},\mu)$ are shown in Fig.\,\ref{fig:FigSM4}(c) to (e).

\subsection{Out-of-plane magnetization}
In the main text, we have analyzed the nonlinearity, $g_3$, and the Kerr coefficient, $K$, only for the case of exchange splittings $(h_{x},h_{y})$ that arise from a magnetization in the plane of the junction. 
An interesting question is how our results change when introducing an exchange splitting, $h_{z}$, which arises from a magnetization that points out of the plane of the junction in the $z$-direction. 

To address this question, we have set $h_{x}=0$ and computed $g_3$ and $K$ as a function of $(h_{z},h_{y})$. Our findings are depicted in Fig.\,\ref{fig:FigSM5}. Our main finding is that the sign reversal in $g_{3}$ and 
in $K$ remains, as shown in Fig.\,\ref{fig:FigSM5}(a) and (b). Interestingly, for the chosen set of system parameters, we find that the $K=0$ condition can be realized at slightly weaker in-plane magnetization if a finite 
out-of-plane component of the magnetization is present. The corresponding resonators frequencies. $\omega_{r}$, and the diode efficiencies, $\eta$, are also shown in Fig.\,\ref{fig:FigSM5}(c) and (d). We also show examples of the Andreev excitation spectrum and the Josephson energy at selected values of $(h_{z},h_{y})$ in Fig.\,\ref{fig:FigSM5}(e). 

\subsection{Double-minimum Josephson energy}
For the interferometer approach, we have already discussed the possibility of a double-minimum Josephson energy. Interestingly, such a double-minimum Josephson energy can also arise in single-junction approach, as highlighted in Fig.\,4 of the main text.

To gain a better understanding of the origin of this double-minimum Josephson energy in the single-junction approach, we have computed several example Andreev excitation spectra, as shown Fig.\,\ref{fig:FigSM6}. 
Assuming that the fermion parity of the junction is not conserved on the measurements time scale, we find that the double-minimum Josephson energy arises when the lowest energy bound state exhibits a zero energy crossing. Hence, we conclude that the origin 
of the double-minimum Josephson energy in the single-junction and in the interferometer approach is different. Indeed, for the interferometer approach no zero-energy crossing of the Andreev bound states was required. However, it is important to emphasize that a zero-energy crossing of Andreev bound states does not guarantee a double-minimum Josephson energy. As is also shown in the example Andreev excitation spectra of Fig.\,\ref{fig:FigSM6}, the zero-energy crossing can also lead to a cusp-like behaviour in the Josephson energy (or, equivalently, a discontinuity in the Josephson current).

\subsection{Mirror-symmetry breaking}
In the main text, we have highlighted that the presence of mirror-symmetry protects $g_{3}=0$ when $h_{y}=0$. Also, we emphasized that 
when the mirror symmetry is broken by local potential variations (arising, for example, from on-site impurities) then $g_{3}\neq0$ when $h_{y}=0$.

We now provide more details on how these results were obtained. Specifically, we have modified the tight-binding Hamiltonian by an additional term,
\begin{equation}
H_{0}+H_{h}\rightarrow H_{0}+H_{h}+\sum_{i,j,s}\varepsilon^{\text{imp}}_{ij} c^{\dagger}_{i,j,s}c_{i,j,s}
\end{equation}
that describes mirror-symmetry breaking on-site impurity potentials. For our simulations, we have chosen $\varepsilon^{\text{imp}}_{03}=0.4,\text{meV}$ and $\varepsilon^{\text{imp}}_{12}=0.3,\text{meV}$, $\varepsilon^{\text{imp}}_{ij}=0$ for all other lattice sites. This impurity configuration is also illustrated in Fig.\,\ref{fig:FigSM7}. We then recomputed the third-order nonlinearity $g_3$, the Kerr coefficient $K$, 
resonator frequency $\omega_{r}$, and the diode efficiency $\eta$ as a function of $(h_{x},h_{y})$. The results are shown both in Fig.\,4 of the main text and in Fig.\,\ref{fig:FigSM8}(a-d). 

Our main finding is that, besides the effective tilting of the phase diagram that leads to $g_{3}\neq0$ when $h_{y}=0$, an additional annulus-like region emerges where the system exhibits a double minimum. However, as shown in Fig.\,\ref{fig:FigSM8}(e), this additional minimum is shallow and appears at much higher energies compared to the global minimum. Therefore, we expect that even with potential variations that break mirror-symmetry, the nonlinearities will still arise from the deep, global minimum of the Josephson energy.

\subsection{Junction length}
In our previous considerations, we have focused on a setup with a junction length of $N_{\text{N}}=7$ lattice sites. An interesting question is if a Kerr-free third-order nonlinearity
can also be reproduced with junctions of shorter length. 

To address this questions, we considered a modified tight-binding model with $N_{\text{N}}=5$ lattice sites in the junction area. Furthermore, we chose $\alpha=25\,\text{meV}$ as the spin-orbit coupling strength, which is still a realistic value for experimental setups involving InAs nanowire Josephson junctions~\cite{matute2022signatures2}. For the chemical potential in the junction region, we chose $\mu_{i,\text{N}}=0.48\,\text{meV}$. All other experimental parameters were left unchanged. Our results are shown in Fig.\,\ref{fig:FigSM9}. 
We find that the Kerr-free third-order nonlinearity is still achievable also for this shorter junction setup. For the chosen parameters, we find that the Kerr-free condition shifts to slightly larger exchange couplings.

\begin{widetext}

\begin{figure*}[!t]
    \centering
    \includegraphics[width=0.9\linewidth]{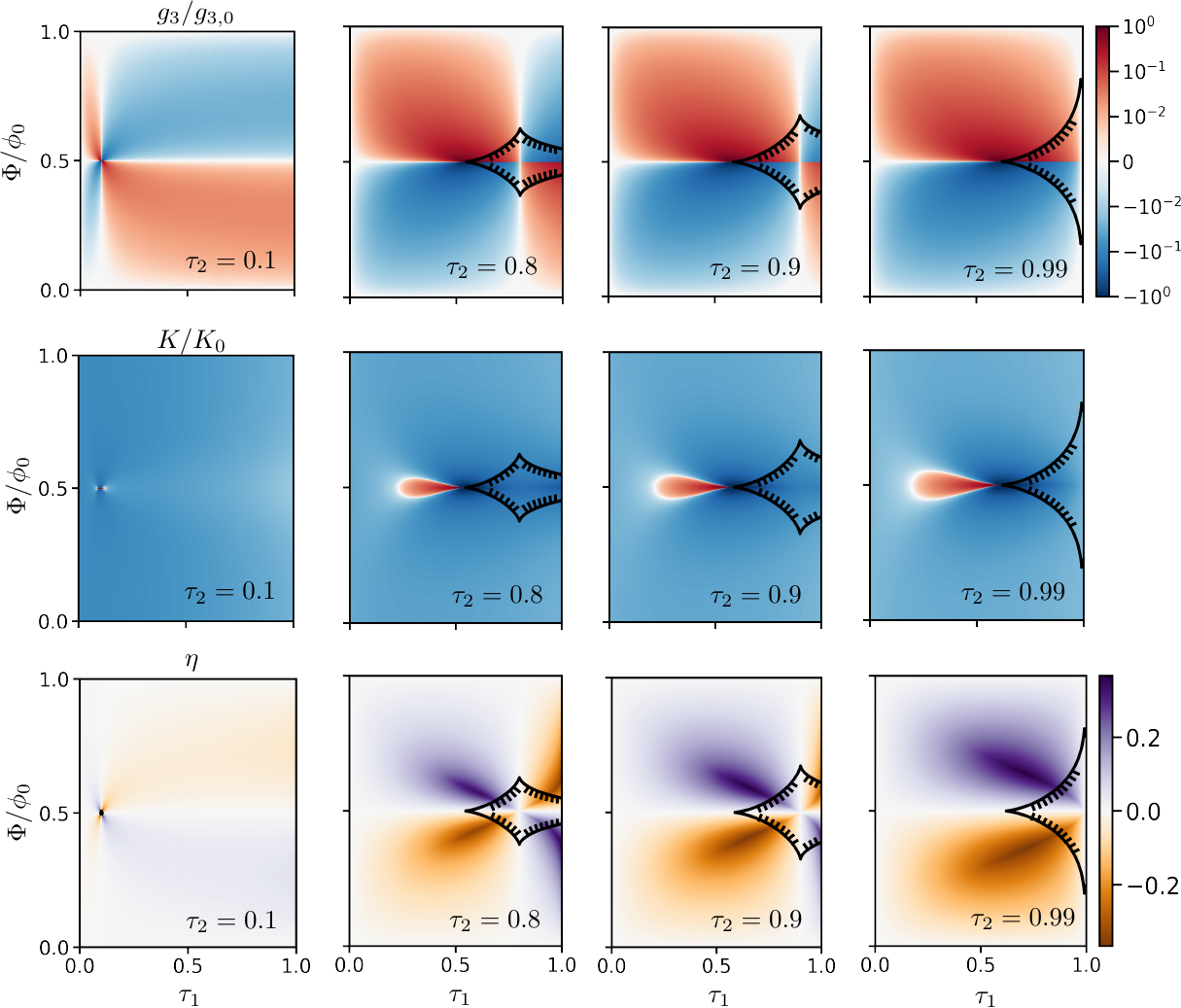}
    \caption{ \textbf{Andreev interferometer properties as a function of flux and transparency for different transparencies of the second junction.}
     Top row shows the third-order nonlinearity, $g_3$, of the interferometer approach versus the transmission, $\tau_{1}$, and the magnetic flux, $\Phi$, for various values of the transmission, $\tau_{2}$. The center and bottom row show the Kerr coefficient, $K$, and the diode efficiency, $\eta$, in the same situations. Notably, the arc length for which the desirable $K=0$ condition can be achieved grows upon increasing $\tau_{2}$.
    }
    \label{fig:FigSM1}
\end{figure*}

\begin{figure*}[!t]
    \centering
    \includegraphics[width=0.9\linewidth]{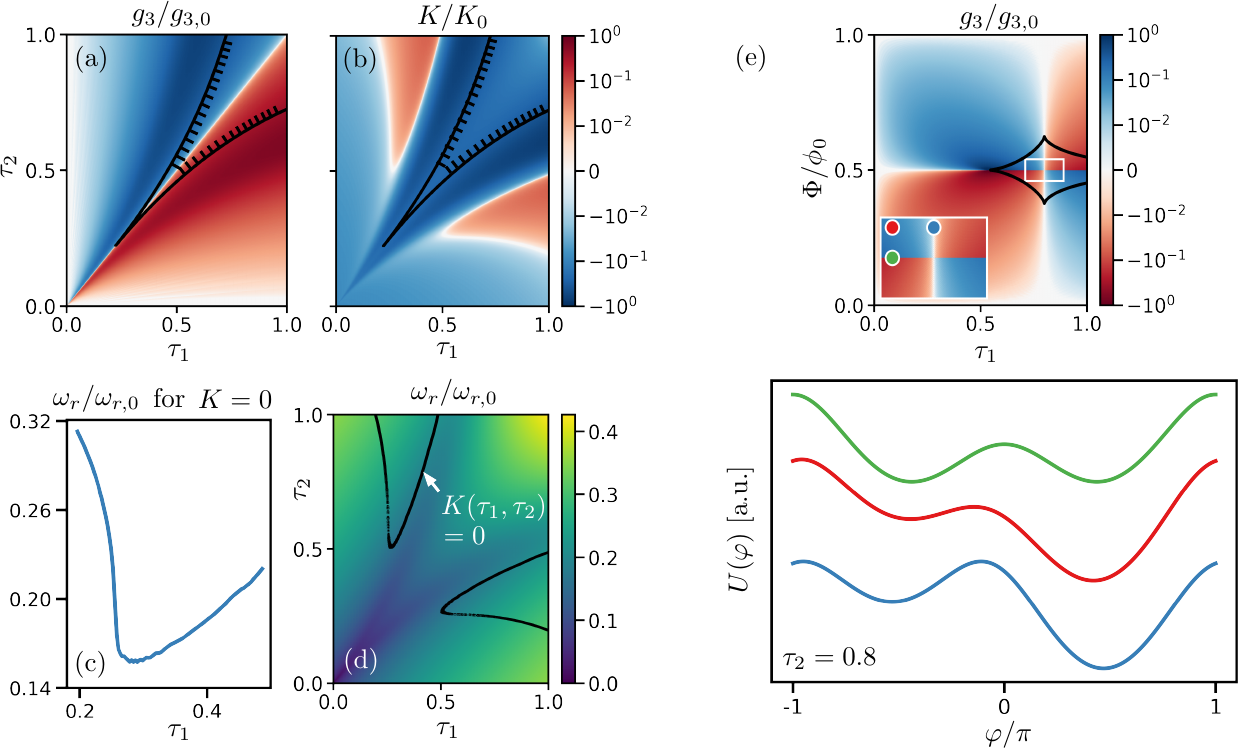}
    \caption{ \textbf{Andreev interformeter properties as a function of both transmissions, and inspection of double-well conditions.}
    (a) Third-order nonlinearity, $g_{3}$, of the interferometer approach versus transmissions, $\tau_{1}$ and $\tau_{2}$ for $\Phi=0.52\Phi_0$. Region bounded by hashed lines marks a double-minimum Josephson potential.  (b) Same as (a) but for the Kerr coefficient, $K$. (c) Resonator frequencies, $\omega_r$, versus $\tau_{1}$ for which the Kerr coefficient vanishes, $K(\tau_{1},\tau_{2}(\tau_{1}))=0$. For any value of $\tau_{1}<0.5$ in the plot, we pick the unique $\tau_{2}=\tau_{2}(\tau_{1})$ such that $K(\tau_{1},\tau_{2}(\tau_{1}))=0$. 
    (d) Resonator frequency, $\omega_{r}$, versus $\tau_{1}$ and $\tau_{2}$. The arcs where $K(\tau_{1},\tau_{2})=0$ are shown as solid black curves. 
    (e) Top: Same as Fig.\,2(a) in the main text but with a zoom-in to the region with a double-minimum Josephson potential. Bottom: Josephson potential, $U$, versus the superconducting phase difference, $\varphi$, for various ($\tau_{1}$,$\Phi$) that are marked with colored dots in the inset of the top plot. 
    For the case of a Josephson potential with two degenerate minima at $\Phi/\phi_{0}=0.5$, we computed the $g_3$ nonlinearity for the minimum within $\varphi\in[0,\pi]$. 
    }
    \label{fig:FigSM2}
\end{figure*}

\begin{figure*}[!t]
    \centering
    \includegraphics[width=0.45\textwidth]{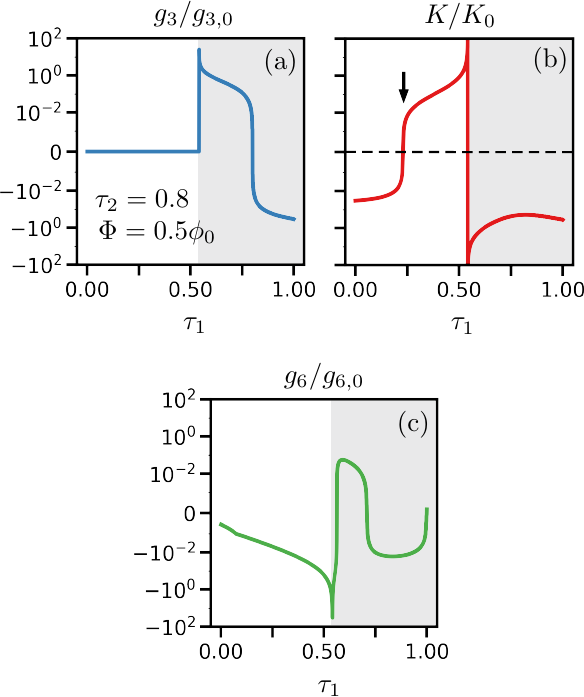}
    \caption{ \textbf{Andreev interferometer higher order nonlinearity inspection.}
    (a) Third-order nonlinearity, $g_3$, for the interferometer approach versus the transmission, $\tau_{1}$,
for fixed $\tau_{2}=0.8$ and $\Phi=0.5\phi_0$ on a symlog scale with threshold $10^{-3}$ and scaling factor $1$. If the Josephson energy, $U(\varphi)$, has a single minimum, then $g_{3}=0$. If it has two degenerate minima, corresponding to the region marked in gray, then $g_{3}\neq0$. In this case, we calculated $g_{3}$ from the mininum within $\varphi\in[0,\pi]$. (b) Same as (a) but for the Kerr coefficient, $K$. The arrow marks the value of $\tau_{1}$ with $g_{3}=K=0$. (c) Same as (a) but for the sixth-order nonlinearity, $g_6$ with $g_{6,0}\equiv [E_{C}/(\hbar\omega_{r})]^{2}/(30\hbar)$. Notably, $g_{6}\neq0$ at the transmission $\tau_{1}$ for which $g_{3}=K=0$.
    }
    \label{fig:FigSM3}
\end{figure*}

\begin{figure*}[!t]
    \centering
    \includegraphics[width=1.05\textwidth]{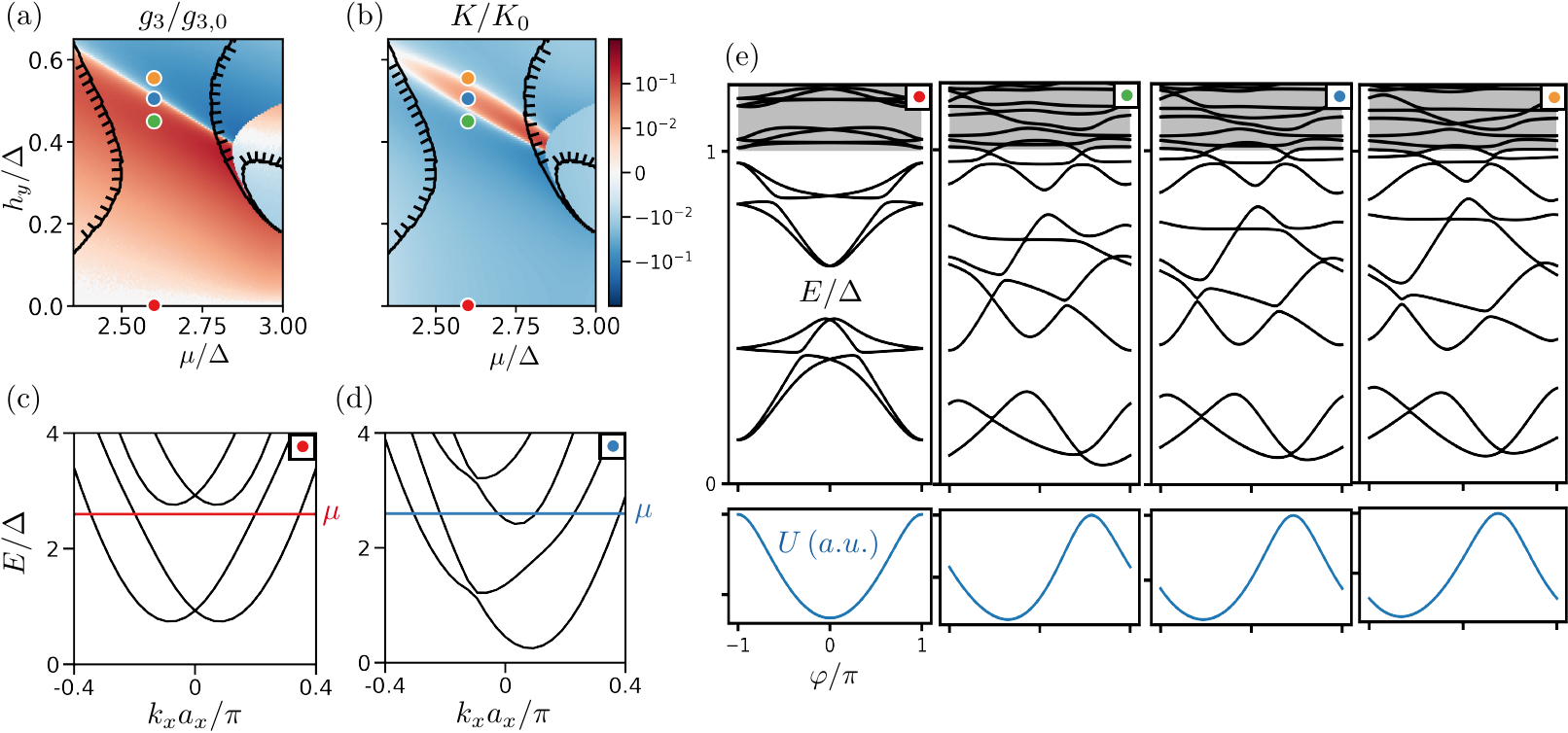}
    \caption{ \textbf{Single-junction properties as a function of exchange energy $h_x$ and chemical potential.}
    (a) Third-order nonlinearity, $g_3$, of the single-junction approach versus the exchange splitting, $h_{y}$ (in the wire plane, perpendicular to the wire), and the chemical potential, $\mu$, in the 
    junction region. Both quantities are normalized by the superconducting gap, $\Delta$.
    (b) Same as (a), but for the Kerr coefficient, $K$. 
    (c-d) Normal state spectrum as a function of the momentum, $k_{x}$ (along the wire), for the parameter values, $(h_{y},\mu)$, highlighted by the red and blue dot in (a-b). The position of the chemical potential, $\mu$, is shown by 
    a red and blue horizontal line. 
    (e) Andreev excitation spectra and Josephson energies, $U$, versus superconducting phase, $\varphi$, for points marked in (a-b).
    }
    \label{fig:FigSM4}
\end{figure*}

\begin{figure*}[!t]
    \centering
    \includegraphics[width=1.05\textwidth]{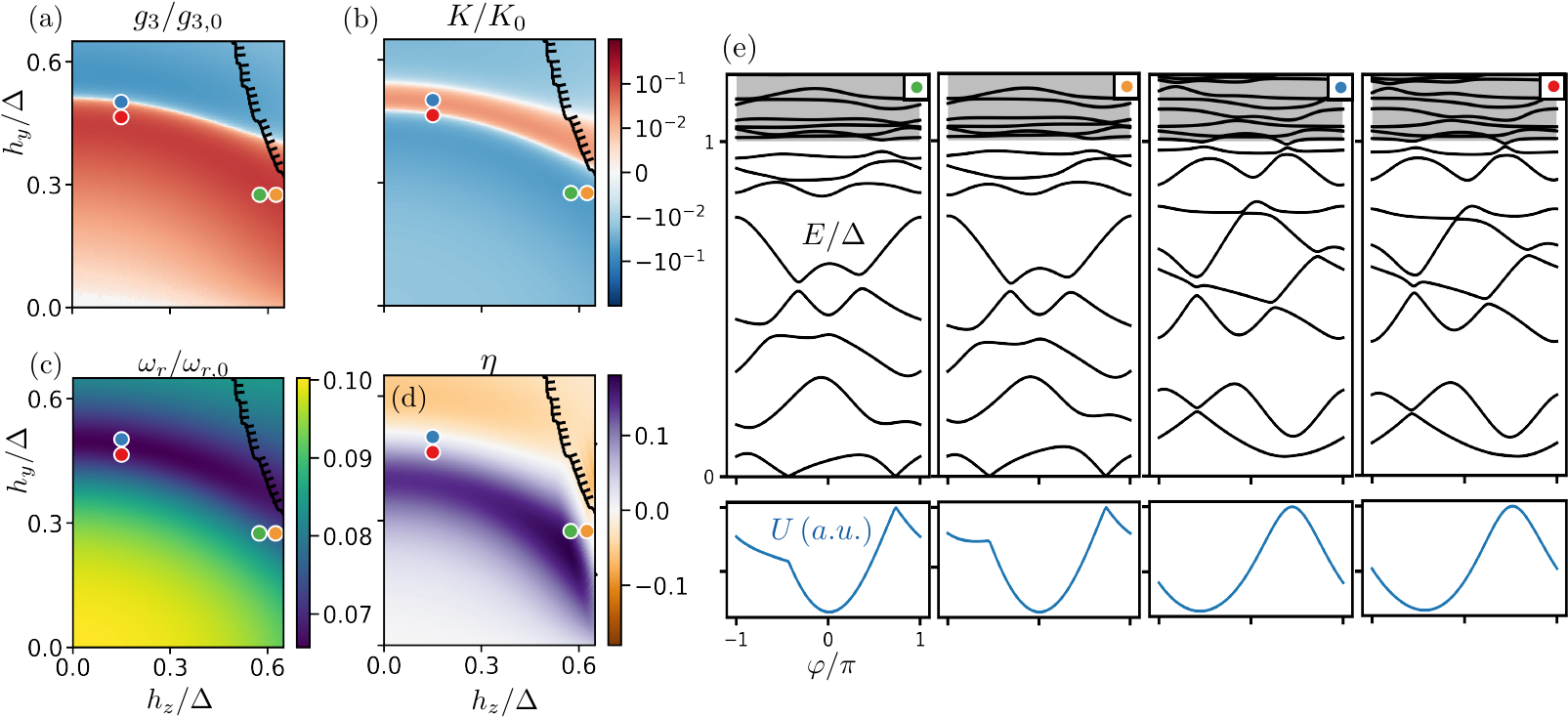}
    \caption{  \textbf{Single-junction properties as a function of exchange energy orientation and magnitude y-z plane.}
    (a) Third-order nonlinearity, $g_3$, of the single-junction approach versus the exchange splittings, $h_{y}$ (in the wire plane, perpendicular to the wire) and $h_{z}$ (out of the wire plane) normalized by the superconducting gap, $\Delta$. (b-d) Same as (a), but for Kerr coefficient $K$, resonator frequency $\omega_{r}$, and diode efficiency $\eta$.
    (e) Andreev excitation spectra and Josephson energies, $U$, versus superconducting phase, $\varphi$, for points marked in (a-d). 
    }
    \label{fig:FigSM5}
\end{figure*}

\begin{figure*}[!t]
    \centering
    \includegraphics[width=1.05\textwidth]{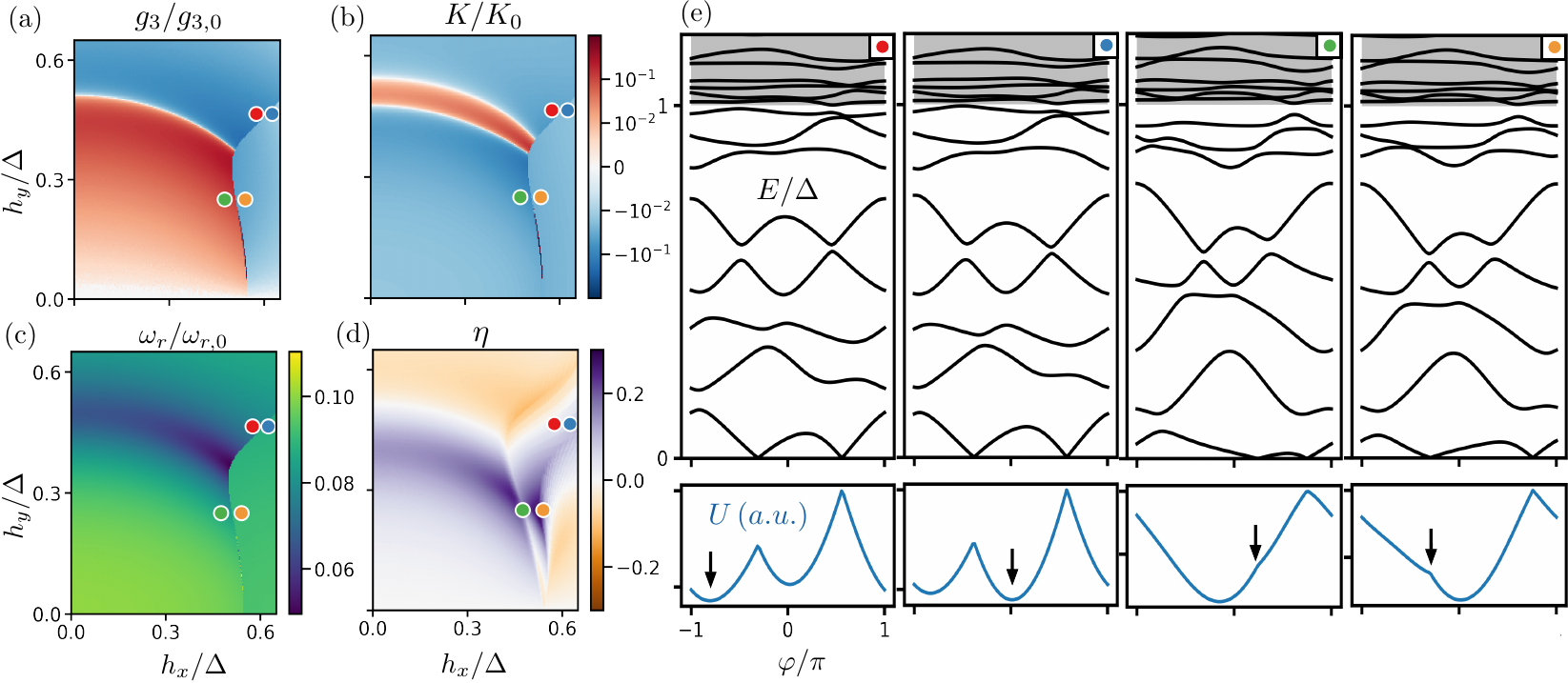}
    \caption{ \textbf{Single-junction properties as a function of exchange energy orientation and magnitude in y-x plane.}
    (a) Third-order nonlinearity, $g_3$, of the single-junction approach versus the exchange splitting, $h_{x}$ (in the wire plane, parallel to the wire) and $h_{y}$ (in the wire plane, perpendicular to the wire) normalized by the superconducting gap, $\Delta$. This plotis a zoom-in  of Fig.\,4(a) in the main text.
    (b-d) Same as (a), but for Kerr coefficient $K$, resonator frequency $\omega_{r}$, and diode efficiency $\eta$. 
    (e) Andreev excitation spectra and Josephson energies, $U$, plotted against the superconducting phase, $\varphi$, for the points indicated in (a-d). In the two left-most figures, a black arrow indicates the lowest minimum for which $g_3$ and $K$ were computed. In the two right-most figures, a black arrow marks a cusp-like feature in the Josephson energy resulting from a zero-energy crossing of the lowest Andreev level.
    }
    \label{fig:FigSM6}
\end{figure*}

\begin{figure*}[!t]
    \centering
    \includegraphics[width=0.87\textwidth]{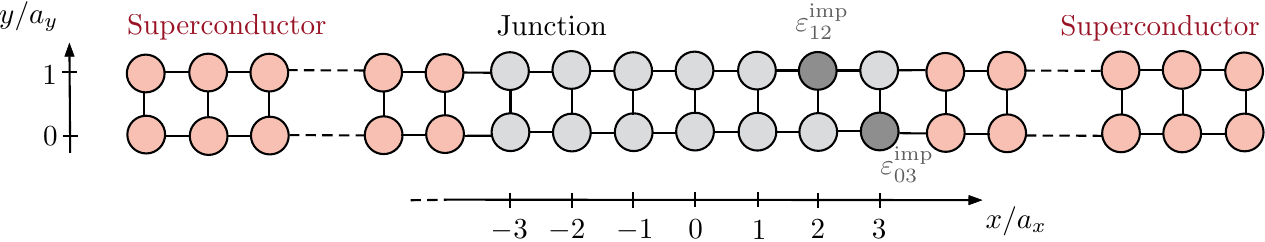}
    \caption{
    \textbf{Schematic of the tight-binding model employed for the single-junction approach simulation.} 
    The superconducting lattice sites are depicted in red, while the normal state sites within the junction are shown in gray. In the scenario of broken mirror symmetry, impurity potentials $\varepsilon^{\text{imp}}_{03}$ and $\varepsilon^{\text{imp}}_{12}$ are located at lattice sites $(x/a_{x},y/a_{y})=(0,3)$ and $(1,2)$, respectively. These lattice sites are also highlighted in dark gray.
    }
    \label{fig:FigSM7}
\end{figure*}

\begin{figure*}[!t]
    \centering
    \includegraphics[width=1.05\textwidth]{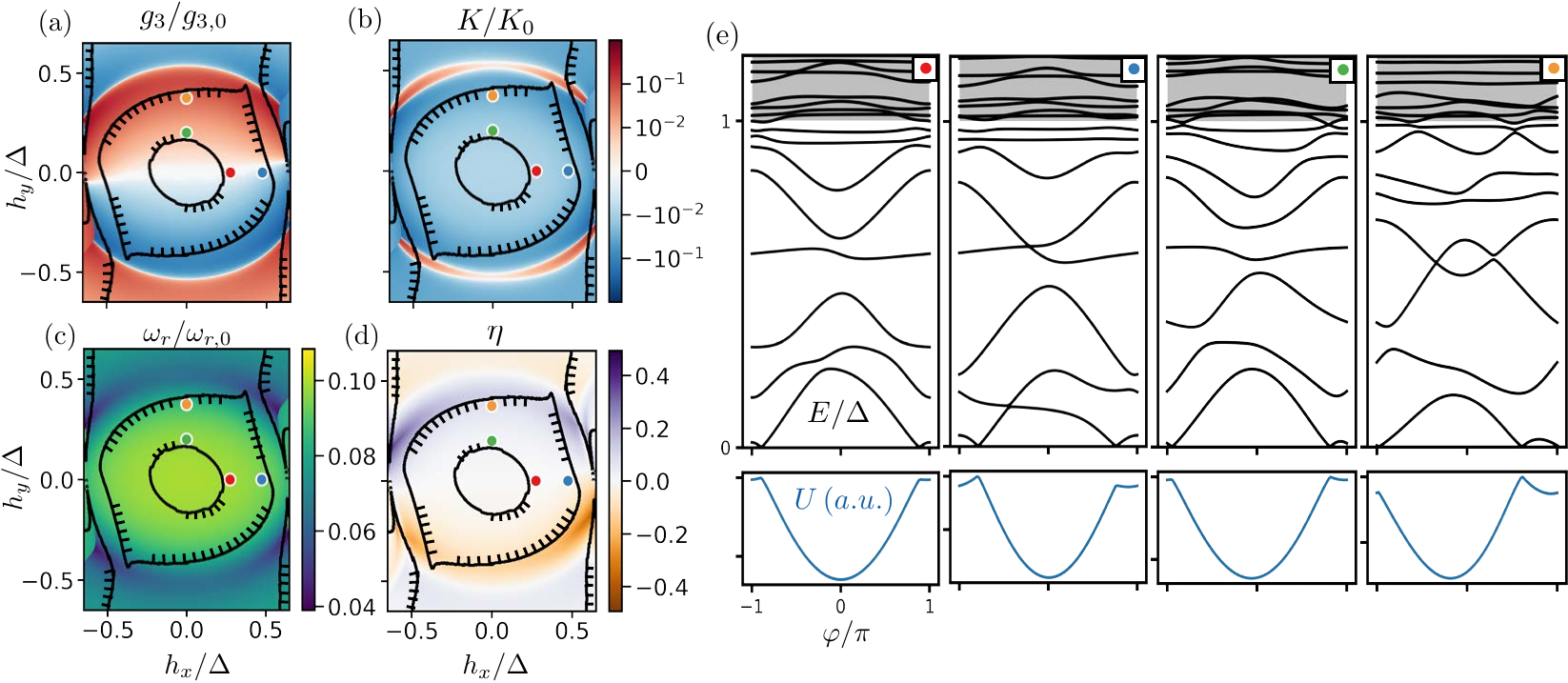}
    \caption{ \textbf{Disordered single-junction properties as a function of exchange energy orientation and magnitude in y-x plane.}
(a-d) Third-order nonlinearity $g_3$, Kerr coefficient $K$, resonator frequency $\omega_r$, and diode efficiency $\eta$ of the single-junction approach as a function of the exchange splittings $h_{x}$ and $h_{y}$, in the presence of local impurity potentials that break mirror symmetry, $\varepsilon^{\text{imp}}_{03}=0.4,\text{meV}$ and $\varepsilon^{\text{imp}}_{12}=0.3,\text{meV}$. Panels (a) and (b) are also included in Fig.\,4 of the main text.
(e) Andreev excitation spectra and Josephson energies $U$ plotted against the superconducting phase $\varphi$ for the indicated points in (a-d).
At these points, an additional minimum emerges in the Josephson energy.
    }
    \label{fig:FigSM8}
\end{figure*}

\begin{figure*}[!t]
    \centering
    \includegraphics[width=1.05\textwidth]{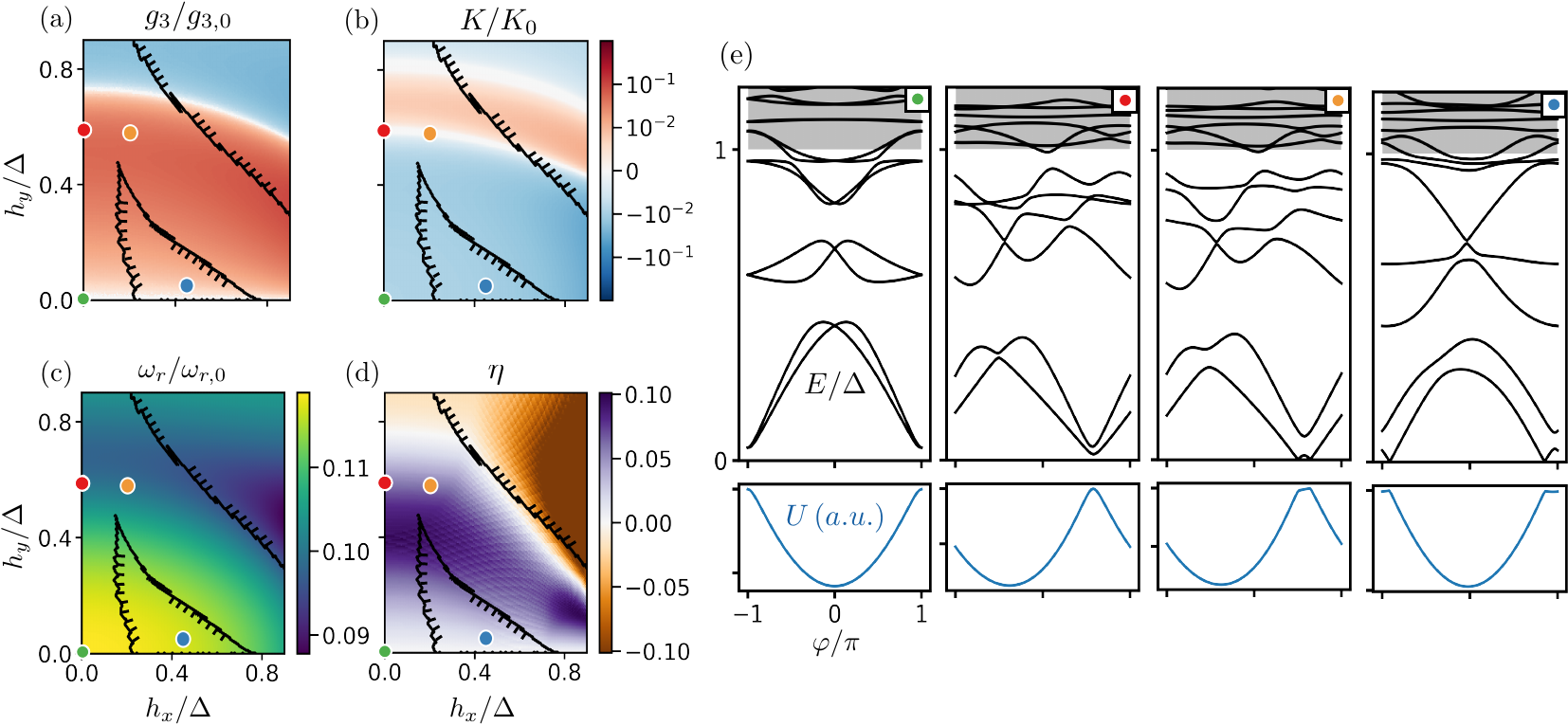}
    \caption{ \textbf{Shorter single-junction properties as a function of exchange energy orientation and magnitude in y-x plane.}
(a-d) Third-order nonlinearity $g_3$, Kerr coefficient $K$, resonator frequency $\omega_r$, and diode efficiency $\eta$ of for a shorter single junction with $N_{\text{N}}=5$ lattice
sites in the junction area. 
(e) Andreev excitation spectra and Josephson energies $U$ plotted against the superconducting phase $\varphi$ for the indicated points in (a-d).
    }
    \label{fig:FigSM9}
\end{figure*}
\end{widetext}

\end{document}